\documentclass[a4paper,11pt]{article}
\pdfoutput=1 

\usepackage{jinstpub} 
\usepackage{verbatim}
\usepackage{graphicx}
\usepackage{subfig}

\title{\boldmath A  Back-End Electronics Based on Fiber Communication  for Small to Medium-Scale Physics Experiments}

\author[a,b]{Jianguo Liu,}
\author[a,b,1]{Yu Wang,\note{Corresponding author.}}
\author[a,b]{Changqing Feng,}
\author[a,b]{Shubin Liu,}
\author[a,b]{Qian Chen,}

\affiliation[a]{State Key Laboratory of Particle Detection and Electronics, University of Science and Technology of China, Hefei, 230026, China}
\affiliation[b]{Department of Modern Physics, University of Science and Technology of China, Hefei, 230026, China}

\emailAdd{wyu0725@ustc.edu.cn}

\abstract{Many small and medium-sized physics experiments are being conducted worldwide. These experiments have similar requirements for readout electronics, especially the back-end electronics. Some experiments need a trigger logic unit(TLU) to provide timing and synchronous control signals. This paper introduces a back-end electronics design for small and medium-sized physics experiments; it adopts a daughter-motherboard structure integrated TLU function to provide greater flexibility. Different interfaces and protocols can be flexibly selected according to data bandwidth requirements. It supports 32 optical fiber interfaces based on a field-programmable gate array (FPGA) of normal IOs with 400Mbps of data bandwidth for each channel. At the same time, it supports 16 high-speed communication interfaces based on GTX port with several Gbps data bandwidth of each channel. For the TLU function, this design has 8 HDMI interfaces and one RJ45 interface to provide synchronous triggers and other control signals, and it has six analog LEMOs and four digital LEMOs to accept asynchronous signals from an external source. These design specifications can meet the needs of most small and medium-sized experiments. This set of back-end electronics has been successfully used in experiments such as PandaX-\uppercase\expandafter{\romannumeral3}, VLAST, and moungraphy. Moreover, it has successfully conducted beam tests with a readout of the data of VLAST detectors at CERN.}


\begin{document}
\maketitle
\flushbottom

\section{Introduction}
\label{sec:intro}
The advancement of science and technology, particularly in electronics-related technologies, has led to the proposal of various back-end electronics solutions for physics experiments. When it comes to small and medium-sized physics experiments, they share similar functional requirements for back-end electronics. These requirements include the transmission rate, bandwidth, and the number of interfaces with the front-end electronics boards\cite{8555999}. A question that arises when designing the back-end electronics module of small to medium-scale experiments is how to develop a set of back-end electronics to satisfy as many small and medium-sized physics experiments as possible. In some experiments, electronic systems require a trigger logic unit (TLU) to provide trigger and control signals to devices employed during test beams\cite{baesso_2019}. It is worth considering how to integrate TLU into the back-end electronics. In addition, it is essential to design a set of portable back-end electronics suitable for preliminary testing, joint debugging, and post-operation of the detector.

Take the ongoing small and medium-sized experiments as examples.  PandaX-\uppercase\expandafter{\romannumeral3} (particle and astrophysical xenon experiment III) searches for the possible neutrinoless double-beta decay with 200kg to one ton of 90\% enriched $Xe^{136}$ in a high-pressure gaseous Xenon TPC \cite{panda_2019}at the world’s most deep underground laboratory CJPL (China Jin Ping Underground Laboratory). It uses 26 front-end boards. The front-end board uses optical fiber to communicate with the back end. The VLAST (Very Large Area Gamma-ray Space Telescope) project was proposed by scientists from the DAMPE (Dark Matter Particle Explorer) cooperation group and aimed to develop a world-leading gamma-ray detection satellite. VLAST will observe gamma rays in the MeV $\sim$ TeV energy range with unprecedented acceptance and hopes to make breakthroughs in research fields such as dark matter, astrophysics, and cosmic rays. The preliminary design of its detector is divided into three parts: anti-coincidence detector (ACD), silicon tracker and low energy gamma-ray detector (STED), and high energy imaging calorimeter (HEIC)\cite{wan2023designoa}\cite{vlast2022}. The prototype was beam tested at CERN on September 12, 2023. The prototype calorimeter composed of crystal particles required four front-end electronics boards. The front end communicates with the back end through the GTX port. TLU is required to provide trigger and control signals. Another experiment, cosmic ray muon imaging, uses eight front-end boards, and the front-end board and back-end are communicated by optical fiber.
\begin{table}[htbp]
\centering
\caption{\label{tab:d} Requirements of the readout system for different experiments.}
\smallskip
\resizebox{\columnwidth}{!}{
\begin{tabular}{c|c|c|c|c}
\hline
Experiments & Trigger rates &Readout channels & Data bandwidth& Trigger mode\\
\hline
PandaX$-$III\cite{panda_2019} & ~10Hz & 10.5K & 102MB/s& Self-trigger\\
\hline
Muography\cite{wangyu} & ~60Hz & 20K-60K & ~10MB/s&Self or external\\
\hline
VLAST \cite{wanqiang}& ~500Hz & 384 & 200MB/s&External\\
\hline
MTPC\cite{chenhaolei} & ~125Hz & 10K & 200MB/s&Self-trigger\\
\hline
\end{tabular}
}
\end{table}

Table \ref{tab:d} shows the readout requirements for different physical experiments. According to the characteristics of these experiments, the readout channels range from hundreds to tens of thousands; the trigger rate ranges from tens to hundreds, and the data bandwidth ranges from a few MB/s to hundreds of MB/s. In the trigger mode, the self-trigger or external trigger is used. The external triggering method uses coincidence detectors to provide trigger information to the electronic system; self-triggering uses multiple instance information provided by the front-end electronics and then sends trigger information to the front-end electronics after being filtered by the back-end. For external triggering, the electronic system needs to be configured with trigger-receiving interfaces.
\\

This paper describes back-end electronics that integrate TLU to meet the practical requirements of small to medium-sized physics experiments like those mentioned above. By building a suitable architecture, the back-end electronics can be applied to as many small and medium-sized physical experiments as possible to save research and development costs and facilitate later maintenance.

\section{Hardware design of the back-end electronics}
The back-end unit uses an inexpensive commercial field-programmable gate array (FPGA) module, offering numerous conventional I/O pins and a limited number of gigabit-per-second transceivers, which can facilitate the design process. The structure of the back-end electronics design is shown in Figure 1. This back-end unit consists of two daughterboards and a motherboard. The daughter boards and the motherboard are connected via Samtec connectors SEAF-50-05.0-L-08-1-A-K-TR\cite{samtec}. The daughterboards are responsible for connecting to front-end cards and other external devices. The motherboard is mainly responsible for data preprocessing and communication with the server or computer.
\begin{figure}[htbp]
\centering 
\includegraphics[width=150mm]{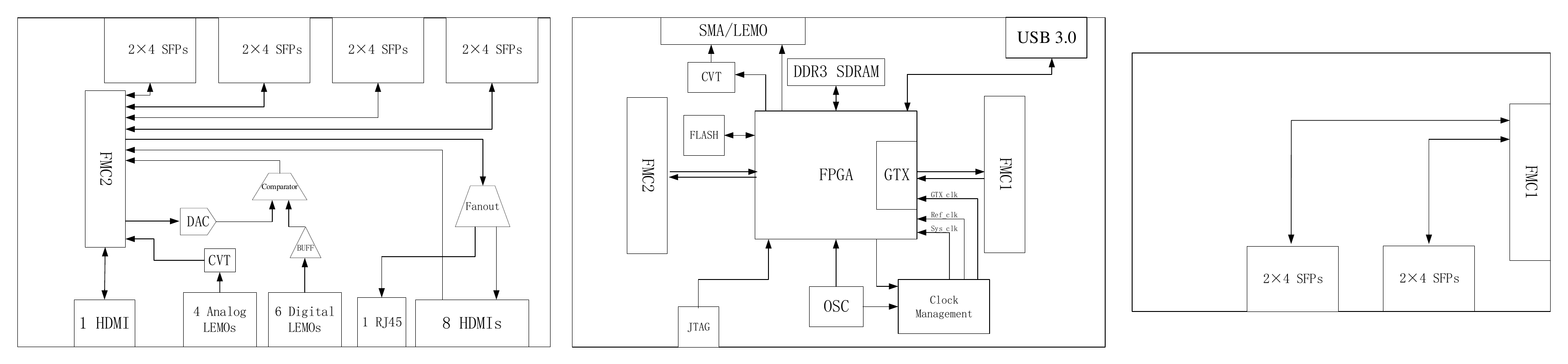}
\caption{\label{fig:j} Schematic of the back-end electronics. }
\end{figure}
In the current design, the two daughterboards have a total of 32 optical fiber interfaces, 16 high-speed optical fibre interfaces, 8 HDMI interfaces, six analog LEMOs, four digital LEMOs, and one RJ45. It is flexible to adjust the interface type and number according to experimental needs. The motherboard has a synchronous dynamic random access memory (SDRAM) chip, a K7 series of FPGAs, and some power and clock management chips. 

\subsection{Structure of clock management}
The clocking scheme is designed in two parts: transmission and generation. In terms of clock transmission, if fiber optic communication is used, the back-end electronics do not require a dedicated link to send clock signals to the front-end electronics. Instead, the clock is sent through the DC-balanced encoding of the data\cite{muography}. The encoded signals are sent to the front-end electronics board via a photoelectric conversion module, whose clock data recovery (CDR) chip decodes the corresponding clock and aligns data from the DC-balanced serial data stream. 

When the front-end electronics sends data to the back-end electronics, its clock and the system clock of the back-end electronics board are the same source, so the data reception and alignment on the back-end electronics can be realized by using the programmable logic resources of the FPGA, and there is no need to design the clock data recovery circuits for each data link. If GTX communication is used, there is a dedicated clock interface on the back-end board that can provide clocks to the front end to make the clock homogeneous.

In terms of clock generation, thie design use a Jitter cleaner chip to meet different clock requirements. The system clock can be generated by the local crystal or Jitter cleaner clock chip, the chip's reference clock is provided by the local crystal or the FPGA.

\subsection{Structure of power management}
The electronics back-end board is powered by a 12V power adapter. The power module uses a quad DC-DC regulator chip LTM4644\cite{ltm4644} to provide operating voltage for the clock chips, FPGA I/O, and optoelectronic conversion devices. Additionally, the low dropout regulator (LDO) TPS7A85\cite{tps7a85} is used to provide operating voltage for the FPGA high-speed transceiver, SDRAM, and other related components.
\subsection{Structure of TLU}
TLU module is responsible for receiving the trigger signal from the outside devices, and the signal will be sent to the FPGA after the threshold comparison to be the signal for the appropriate spreading and conformity through the HDMI, RJ45 interface output control signals.
\begin{table}[htbp]
\centering
\caption{\label{tab:a} HDMI pin map.}
\smallskip
\begin{tabular}{c|c|c|c|c}
\hline
HDMI pin & TLU signal & &HDMI pin & TLU signal\\
\hline
1 & CLK\_p & &11 & GND\\
\hline
2& GND &&12&SPARE\_n\\
\hline
3&CLK\_n&&13&NC\\
\hline
4&TRIG-ID\_p&&14&NC\\
\hline
5&GND&&15&TRIG\_p\\
\hline
6&TRIG-ID\_n&&16&TRIG\_n\\
\hline
7&BUSY\_p&&17&GND\\
\hline
8&GND&&18&NC\\
\hline
9&BUSY\_n&&19&NC\\
\hline
10&SPARE\_p&&&\\
\hline
\end{tabular}
\end{table}

\begin{table}[htbp]
\centering
\caption{\label{tab:b} RJ45 pin map.}
\smallskip
\begin{tabular}{c|c|c|c|c}
\hline
RJ45 pin & TLU signal & &RJ45 pin & TLU signal\\
\hline
1&CLK\_n&&5&TRIG-ID\_p\\
\hline
2&CLK\_p&&6&BUSY\_p\\
\hline
3&BUSY\_n&&7&TRIG\_n\\
\hline
4&TRIG-ID\_n&&8&TRIG\_p\\
\hline
\end{tabular}
\end{table}

The interface of the trigger module is EPY.00.250.NTN LEMO\cite{lemo}, of which there are six analog LEMOs and four digital LEMOs. Analog LEMO input signal through the inverting amplifier LMH6722\cite{lmh6722} connected to the positive end of the comparator AD8564\cite{ad8564}; the negative end of the comparator connected to the operation amplifier and DAC, the main role of the op-amp is to improve the ability of the DAC to carry loads, the FPGA can control the DAC to set the threshold. the DAC model is  AD5391\cite{ad5391}, operation amplifier model is AD8544\cite{ad8544}. As shown in table \ref{tab:a} and table \ref{tab:b}. The control signals include CLK, BUSY, TRIG, TRIG-ID, and SPARE, which are received or sent by the FPGA. The TLU module has a total of 8 HDMI interfaces and one RJ45 interface. Considering the number of pins and topology space limitations,  in the HDMI interface,  besides the BUSY signal, the remaining four signals are generated by differential fan-out chip  DS90LV110T\cite{ds90lv110}. The RJ45 interface signals connect with FPGA through the bidirectional buffer; the users can customize the relevant signals as input signals or output signals, so the RJ45 communication protocols have a certain degree of flexibility. The motherboard provides the 5V power supply to the TLU module via the Semtec connector. The power supply to the on-board DACs, inverting amplifiers, and other devices needs to be provided by the LDOs and the reference power supply chip.\\

\begin{figure}[htbp]
\centering 
\subfloat[The front view]{\label{fig:subfig1}\includegraphics[width=70mm]{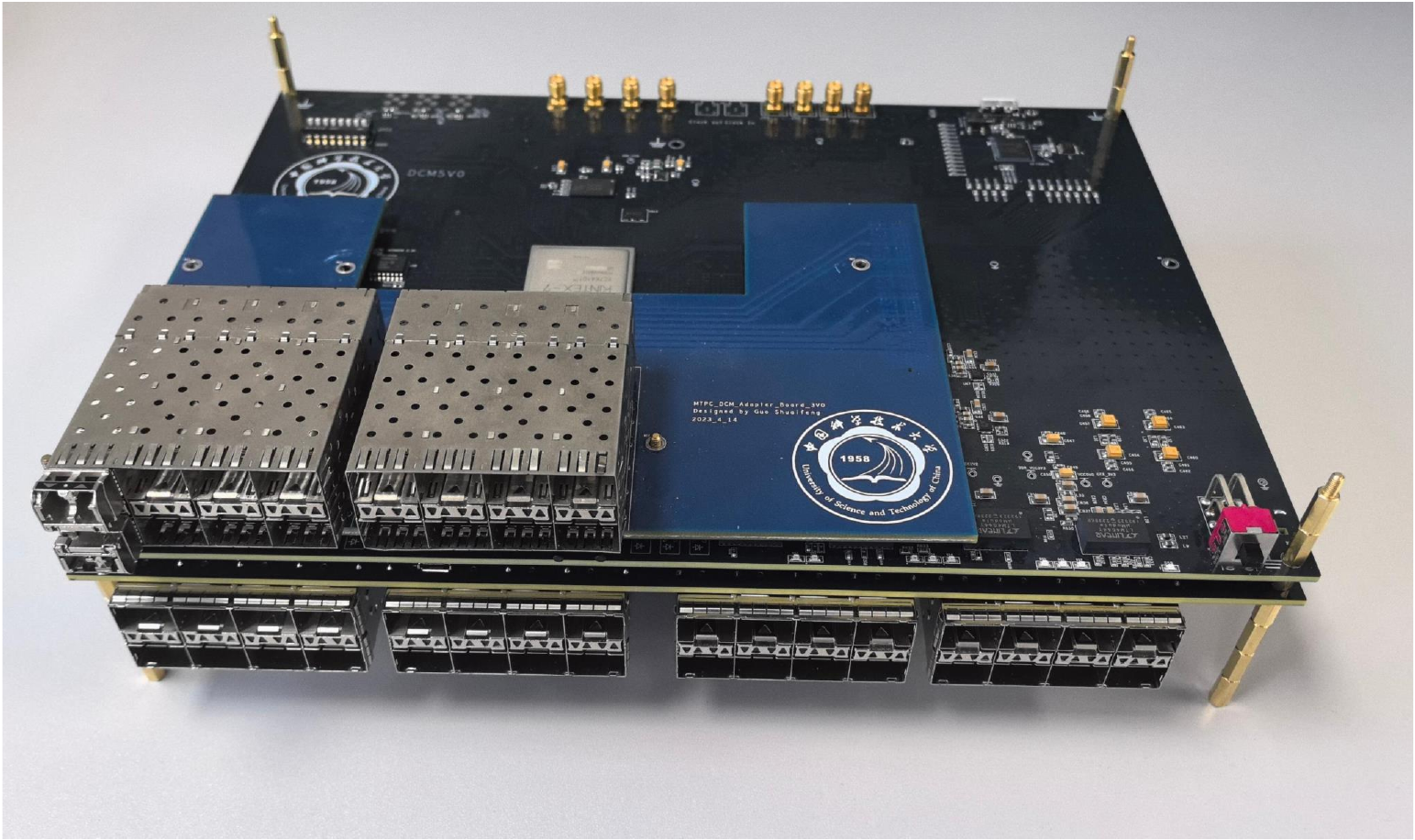}}
\subfloat[The dorsal view]{\label{fig:subfig2}\includegraphics[width =70mm]{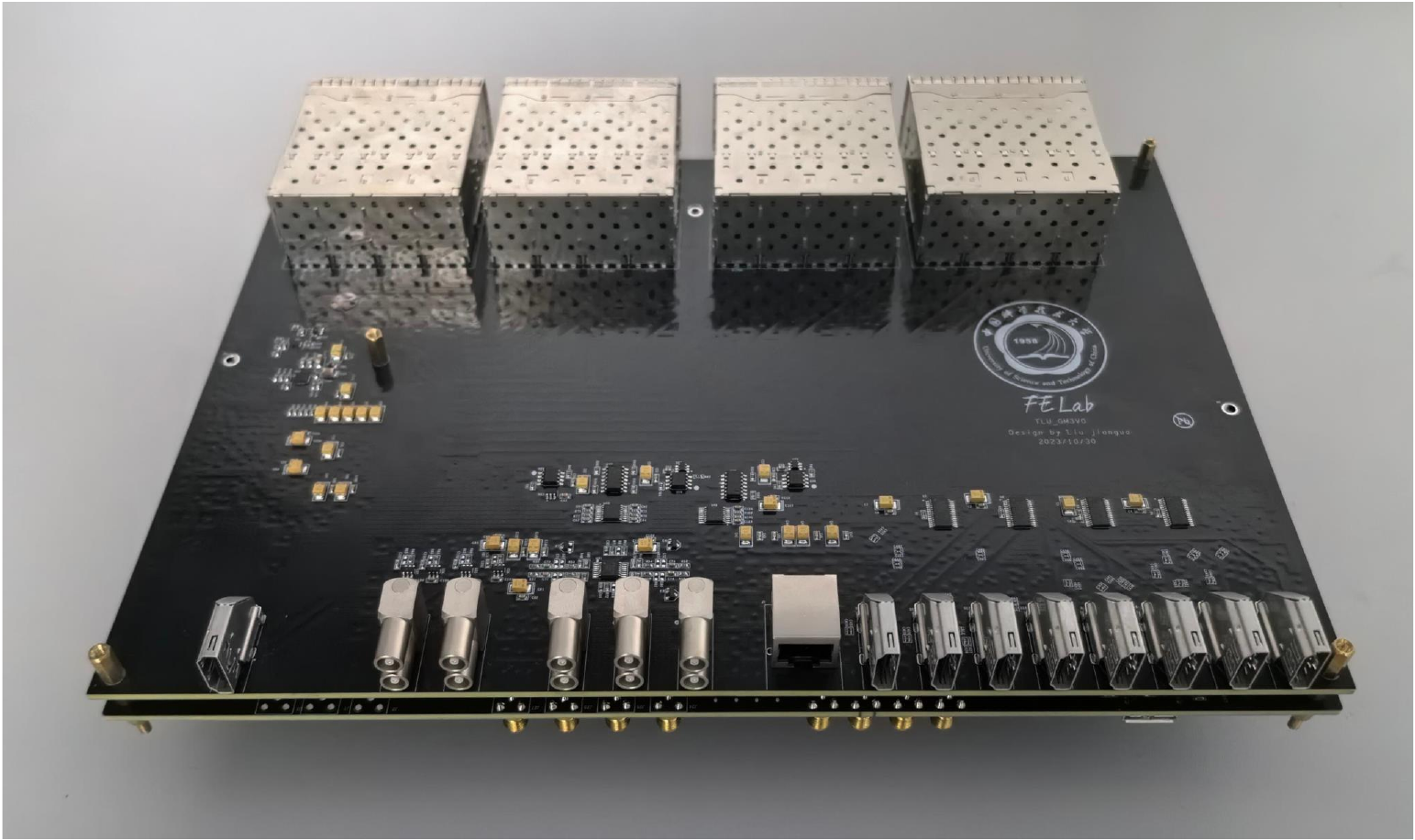}}
\caption{\label{fig:i} Picture of the back-end electronics. }
\end{figure}

The specific photos are shown in Figure\ref{fig:i}.In the front view, the blue board is a daughterboard with eighteen high-speed communication interfaces, and the black middle board is the motherboard. In the dorsal view, the upper blackboard is the daughterboard with TLU and fiber communication interfaces. 

\section{Function design of the back-end electronics}
The main logic functions of back-end electronics include data link coding, data receiving and filtering, trigger receiving and generation, USB chip control, high-speed serial interface control, etc. The functional design of each module is described below.
\subsection{Design of data transmission protocol}
The physical link bandwidth is split into three channels to enhance the data transmission protocol's efficiency: data channel, trigger channel, and configuration channel. The data channel is used for acquiring detector data. The trigger channel is used to transmit the synchronous trigger signal. The configuration channel is used to send control commands. Moreover, two types of data transmission protocols are used to meet the bandwidth requirements of different experiments. 

Based on the optical fibre link implemented by normal IO, the corresponding communication protocol can be flexibly designed according to experimental needs; for front-end electronics to decouple the system clock from the downlink, DC-balanced encoding is required. Since the amount of downlink data is small, Manchester encoding\cite{manchester}, which is the easiest to implement, is used as the downlink encoding scheme. The basic principle of Manchester encoding is to use a low-level to high-level transition to represent a logic 0 and a high-level to low-level transition to represent a logic 1. In the design, the downlink data rate is 100Mbps, which is divided into 3 data channels using time-division multiplexing to transmit trigger signals, configuration commands, and downlink data packets, respectively. The trigger signal bandwidth is 50Mbps, and the configuration command and downlink data packets each occupy 25Mbps bandwidth. The uplink needs to transmit a large amount of collected data. In order to avoid losing the transmission bandwidth, self-synchronous scrambling\cite{gilee1994} is used to implement data encoding. When the amount of data is large enough, the DC balance of the serial data stream can be ensured by using the scrambler method. The uplink data rate is designed to be 400Mbps, divided into three channels. The trigger channel and command channel bandwidth are 100Mbps, and the data channel bandwidth is 200Mbps. 

The uplink data is synchronized with the back-end electronics board clock, so there is no need to design corresponding circuits on the back-end electronics board to recover its clock. In order to avoid metastability in the data link, dedicated hardware resources provided by the FPGA need to be used to appropriately delay the input pin signal. Xilinx's k7 series FPGA integrates a dedicated data input delay adjustment module IDELAYE2, which can independently delay the input signal of each pin\cite{product}. In this design, the single adjustment step size is set to 78ps, and the delay adjustment time range is 0ns $\sim$ 2.5ns, which meets the setup and hold time required by Kintex-7 FPGA.
\begin{figure}[htbp]
\centering 
\includegraphics[width =140mm]{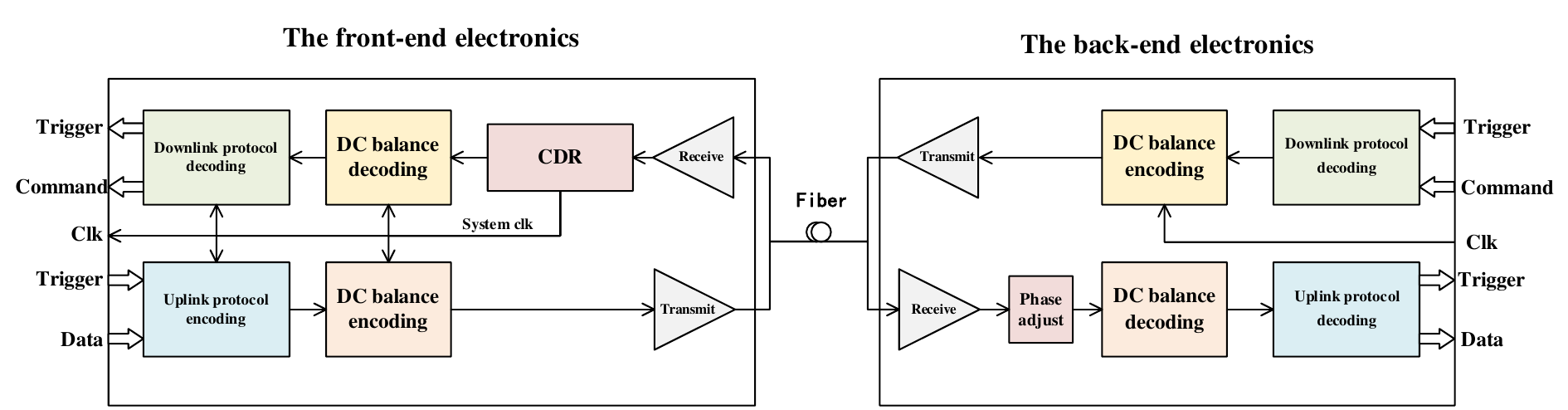}
\caption{\label{fig:w} Diagram of fiber protocol. }
\end{figure}

For experiments that require more data bandwidth, normal I/O optical fiber communication cannot meet the requirements, and high-speed communication protocols are required. The data transmission interface with the front-end electronics is placed on the daughter board, and the data transmission can be carried out with the front-end electronics in different experiments by developing the daughterboard based on different interfaces, which saves the cost of replacing the motherboard and makes the back end have strong versatility. In this design, 16 optical fiber transceiver interfaces are placed on the daughterboard using the GTX communication protocol\cite{gtp}. The bit width of the GTX protocol is 16 bits, including 8 bits for the data channel, 4 bits for the command channel, and 4 bits for the trigger channel. The data rate of each transceiver interface is 2.4 Gbps, the data rate can be adjusted according to the experimental requirements, and the data is connected to the FPGA high-speed interface through high-density connectors.

GT high-speed transceiver interface is divided into two relatively independent parts: transmit and receive. The sending part is mainly used for 8B/10B encoding and parallel series conversion for parallel data. The receiving part is used to serialize, convert, align, and decode the input high-speed serial data. The whole sending process is realized by two parts: TX PCS and TX PMA. PCS mainly completes the parallel data 8B/10B encoding. PMA is mainly used for parallel series conversion of data, and according to the high-frequency clock generated by the GT interface PLL, the serial data is controlled, and the input clock of the PLL is the optical fiber interface reference clock. In the PCS part, the design only uses the 8B/10B encoding function to process the data to ensure that the transmission process has a definite delay. The fiber reference clock is 120 MHz, and since the serial data rate of 2.4 Gbps is an integer multiple, the phase uncertainty caused by the PLL division of the GT interface is avoided so that the delay of the entire transmission process is fixed. The receiving part has a similar structure to the transmitting part, but it is quite different from the PMA part. When serial data is received, the CDR circuit generates a clock for deserializing the data, and the clock is used to process and control the data after serial and conversion after frequency division.

\subsection{Design of trigger}
In this design, the trigger modes of the back-end electronics include self-trigger, external trigger, and test trigger. The back-end electronics also integrates a TLU module. 

In the self-triggering mode, the back-end electronics extract the trigger signal from the uplink data link, determine the number of triggered front-end electronics, and then issue the trigger number and trigger command to each front-end electronics board. In the external trigger mode, the trigger signal is input to the FPGA through the SMA (SubMiniature version A) or HDMI interface on the back-end electronics board. After the trigger module receives the external trigger, it sends the trigger number and trigger command to the front-end electronics board.  In the test trigger mode, the back-end electronics uniformly send trigger signals to the front end. The frequency of the trigger signal is generally from tens of hertz to thousands of hertz and can be adjusted according to needs. This mode is generally used to test whether the electronic system works properly and collect baseline signals.

The TLU module in this design provides timing and synchronous control signals to test-beam readout electronics. It accepts the asynchronous trigger signals from an external source, such as beam-scintillators, and generates synchronous triggers or other control signals\cite{baesso_2019}. It also accepts busy signals or other veto signals from DUTs.
\begin{figure}[htbp]
\centering 

\subfloat[HDMI timing diagram]{ \label{fig:subfig1}\includegraphics[width=75mm]{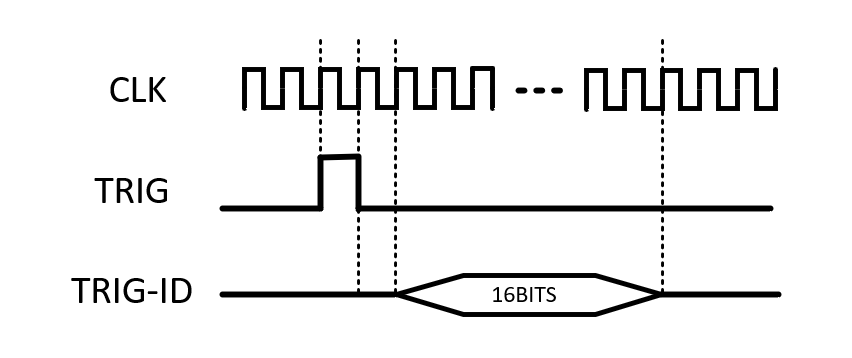}}

\subfloat[RJ45 timing diagram]
{ \label{fig:subfig2}\includegraphics[width=70mm]{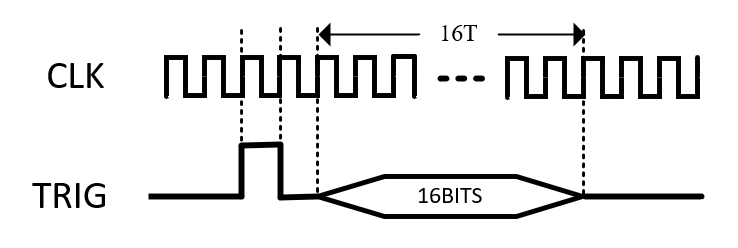}}
\caption{\label{fig:x} Schematic of HDMI and RJ45 timing diagram. }
\end{figure}
The TLU module has a total of 8 HDMI interfaces and one RJ45 interface. Five pairs of signals are set for the HDMI interface: CLK, TRIG, TRIG-ID, BUSY, and SPARE. CLK is a 40M clock signal, and other control signals are synchronized with this. The SPARE signal can be customized according to needs. In the current design, this signal is idle. Compared with HDMI, the RJ45 interface only has four pairs of signals: CLK, TRIG, BUSY, and CONT. CLK is a 40M clock signal, and the other signals are synchronized with this. The trigger signal and trigger count signal are output through TRIG. The HDMI and RJ45 interface signal timing diagram is shown in the figure \ref{fig:x}.

\subsection{Design of USB control}
Through firmware settings, the USB chip can be configured into Slave FIFO (First-In First-Out) mode\cite{usb}, and the FPGA serves as the master device to control the USB chip. In this mode, the chip storage space is configured as two terminals (Endpoints), which are responsible for data upload and delivery, respectively. Each terminal is configured with a certain number of FIFOs and has corresponding FIFO status flags connected to the FPGA.
\begin{figure}[htbp]
\centering 
\includegraphics[scale =0.5]{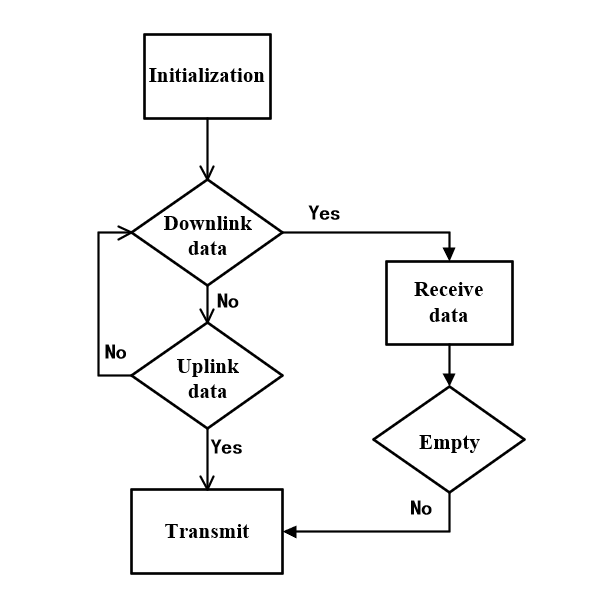}
\caption{\label{fig:u} diagram of USB control. }
\end{figure}

In order to ensure that the host computer controls the readout electronics, the USB chip readout control module is in downstream priority mode. The state machine keeps detecting the status of the downstream FIFO. After the host computer sends a command to the USB chip, the downstream FIFO becomes non-empty. After detecting this status, the FPGA reads the data from the downstream FIFO and then sends the data to the command parsing module for processing. When there is data to be uploaded, the FPGA writes data to the FIFO of the upload terminal through the address control line and data bus. When the current FIFO is full, it will automatically switch to the next FIFO and upload the data of the current FIFO to the computer or server. The host computer accesses the corresponding memory space to obtain the data uploaded by the FPGA. The working clock of the USB chip is 100MHz, and its data bus width is 32 bits, so the theoretical data rate is 400MB/s.

\section{Test and application of this back-end electronics}
This set of back-end electronics has been successfully used in experiments such as the Pands-XIII experiment, the cosmic ray muon imaging experiment, and VLAST. This chapter will introduce the working of the back-end board in these experiments. In addition, this section will also introduce the test of the back-end electronics. The test content includes interface performance testing and TLU function test.
\subsection{Test of bit error rate}
The front-end electronic board and the back-end electronic board communicate through a custom optical fiber protocol or GTX protocol. In order to verify the performance of the interface and protocol, a bit error rate test is carried out for the normal IO optical fiber interface and GTX interface.

For normal IO optical fiber interfaces, use the polynomial $\rm X^{31} + X^{28} + 1 $\cite{prbs}as the pseudo-random number sequence to generate the polynomial. The pseudo-random number sequence generated by the back-end electronics board is sent to the front-end electronics board through the downlink. The front-end electronics board verifies the input data and counts the number of verified bit errors. At the same time, pseudo-random number sequences are also generated in the front-end electronics to test the uplink. Use the VIO function provided by Xilinx FPGA to control the generation and stop of the PRBS test and monitor the number of transmission errors through VIO.

As mentioned in Section 3.1, The bandwidth of the downlink is 200Mbps, and the bandwidth of the uplink is 400Mbps. The test lasted for 208 hours in total. The downlink transmitted a total of $1.498\times10^{14}$bits of data, and the uplink transmitted a total of $2.995\times10^{14}$ bits of data. No bit errors were detected. See \eqref{eq:m}.
\begin{equation}
\label{eq:m}
\begin{split}
 \rm  CL = 1 - e^{-N \times BER_S} 
      \times \sum_{k=0}^E \frac{(N \times BER_S)^k}{k!}
\end{split}
\end{equation}
CL represents the BER confidence level, and N represents the number of bits transmitted, E represents the number of bit errors detected. At 95\% confidence level, the downlink BER is less than $2\times10^{-14}$, and the uplink BER is less than $ 1\times10^{-14}$.

For GTX interfaces, use the IBERT function provided by Xilinx FPGA to generate the PRBS and monitor the error bits. Communication bandwidth is 2.4 Gbps, tested for 50 hours, with a total transmission number of $\rm4.3\times10^{14}$ bits. At 95\% confidence level, the BER is less than $\rm7\times10^{-15}$.

\subsection{Function test of TLU}
   The TLU accepts Pulse signals from external sources, such as scintillators, and generates a global trigger and some Synchronous control signals. In order to verify that TLU can work normally and stably, functional testing on the TLU module is necessary. A signal source is used to generate a pulse signal, and the signal is sent to the LEMO interface of the TLU module. An oscilloscope is used to detect the signals in the HDMI and RJ45 interfaces. The upper limit of the acceptable trigger rate of TLU is about seveal MHz.Further adjustments to the triggering logic can further increase this upper limit, but for most small and medium-sized experiments, the trigger rate at 1 MHz is completely sufficient. We conducted some tests on common trigger rates, and the test results are shown in Table\ref{tab:c}.

 The signal source was used to test signals with different trigger rates, and there was no error in trigger counting. This TLU system was used with a plastic scintillator to conduct cosmic ray muon tests on the VLAST sub-detector HEIC. During nearly a week of testing, the TLU was able to work normally and stably.
\begin{table}[htbp]
\centering
\caption{\label{tab:c} Trigger test of different rates.}
\smallskip
\begin{tabular}{|c|c|c|c|}
\hline
test rates(Hz) & test times & error times & successful rates(\%)\\
\hline
100 & 20k & 0 & 100\\
\hline
1k & 200k & 0 & 100\\
\hline
10k & 800k & 0 & 100\\
\hline
100k & 6.5M & 0 & 100\\
\hline
1M & 20M &0&100\\
\hline
\end{tabular}
\end{table}

\subsection{Apply in  PandaX-\uppercase\expandafter{\romannumeral3}}
\begin{figure}[htbp]
\centering 
\includegraphics[scale =0.4]{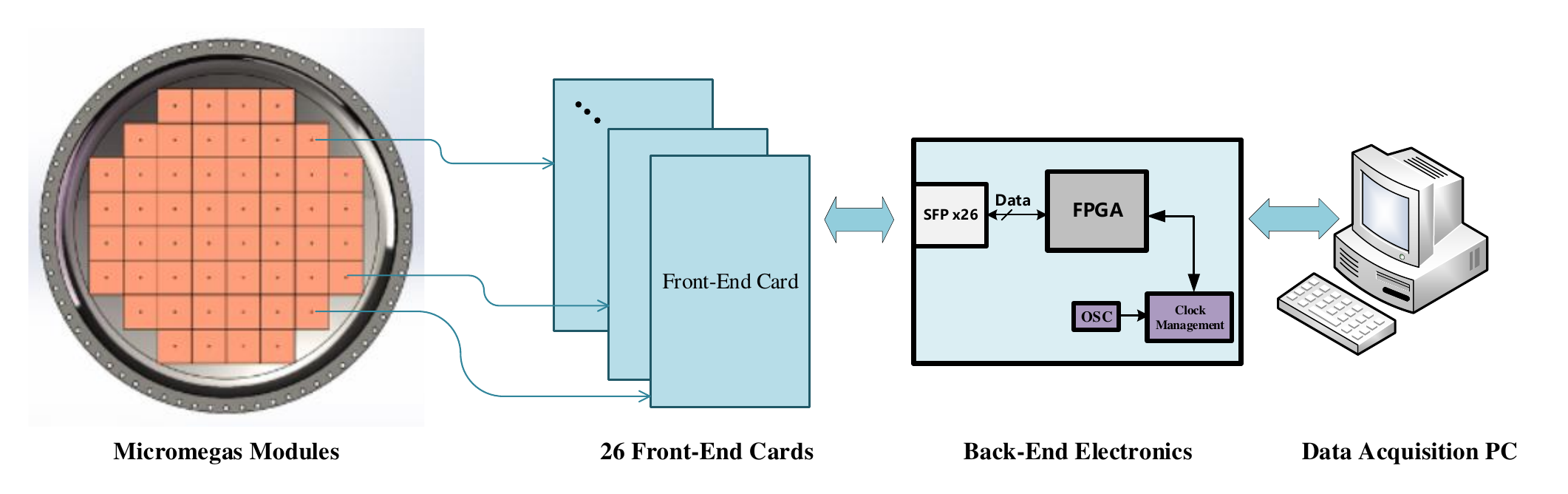}
\caption{\label{fig:y} Schematic of electronics test in PandaX-\uppercase\expandafter{\romannumeral3}. }
\end{figure}
The  PandaX-\uppercase\expandafter{\romannumeral3} project detector has yet to be officially operated, so the signal generator is used instead of the cosmic ray signal in the electronics test. As shown i  figure \ref{fig:y}, the back-end electronics are connected to 26 front-end boards through normal IO optical fiber interfaces. According to the  PandaX-\uppercase\expandafter{\romannumeral3} experimental requirements, the front-end ASIC chip self-scale test, integral nonlinear test of the front-end board channels, baseline (noise) acquisition test, and data acquisition test were performed.
\begin{figure}[htbp]
\centering 
\subfloat[ test of noise]
  {
      \label{fig:subfig1}\includegraphics[width =55mm]{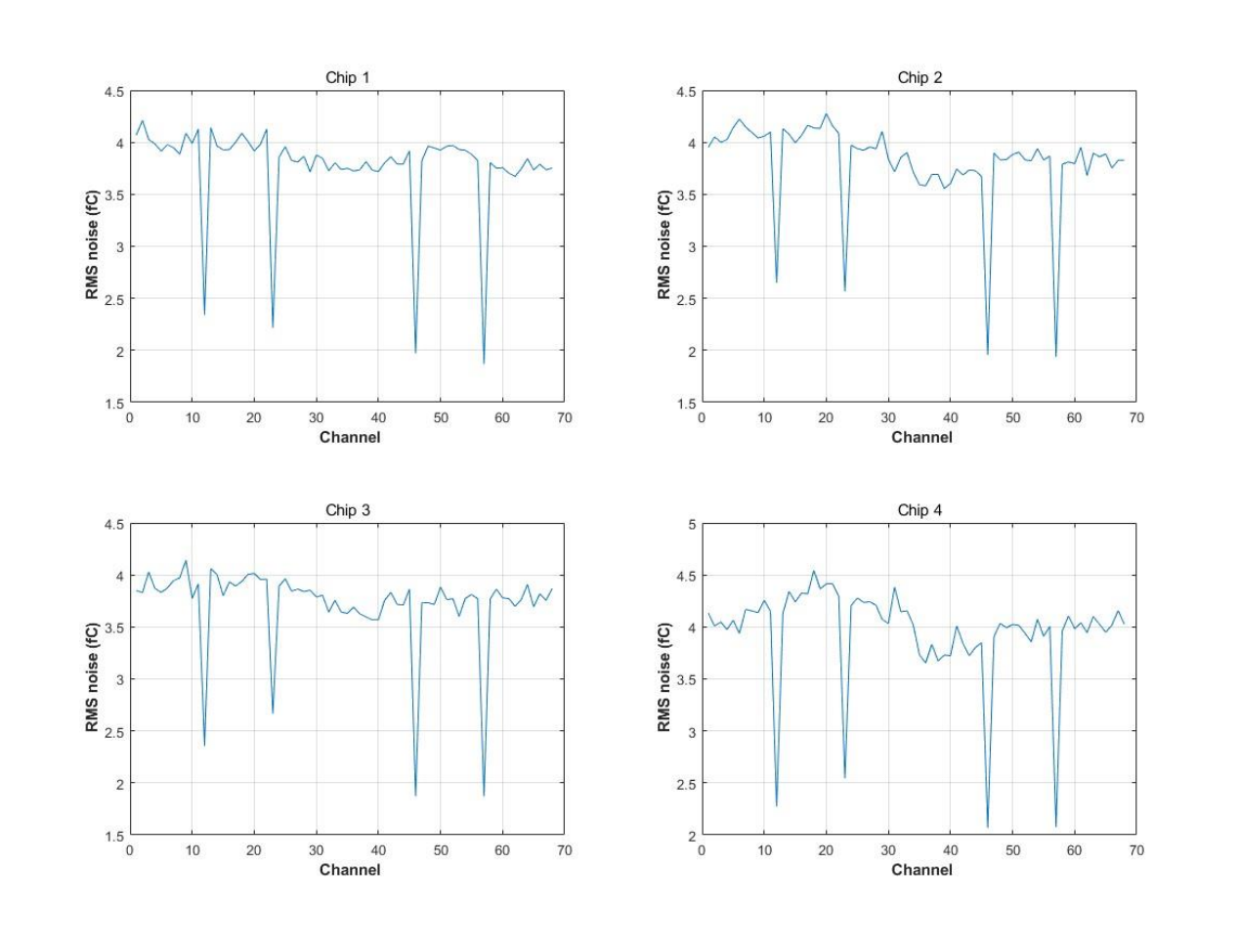}
  }
  \subfloat[beam test of self-calibration]
  {
      \label{fig:subfig2}\includegraphics[width =55mm]{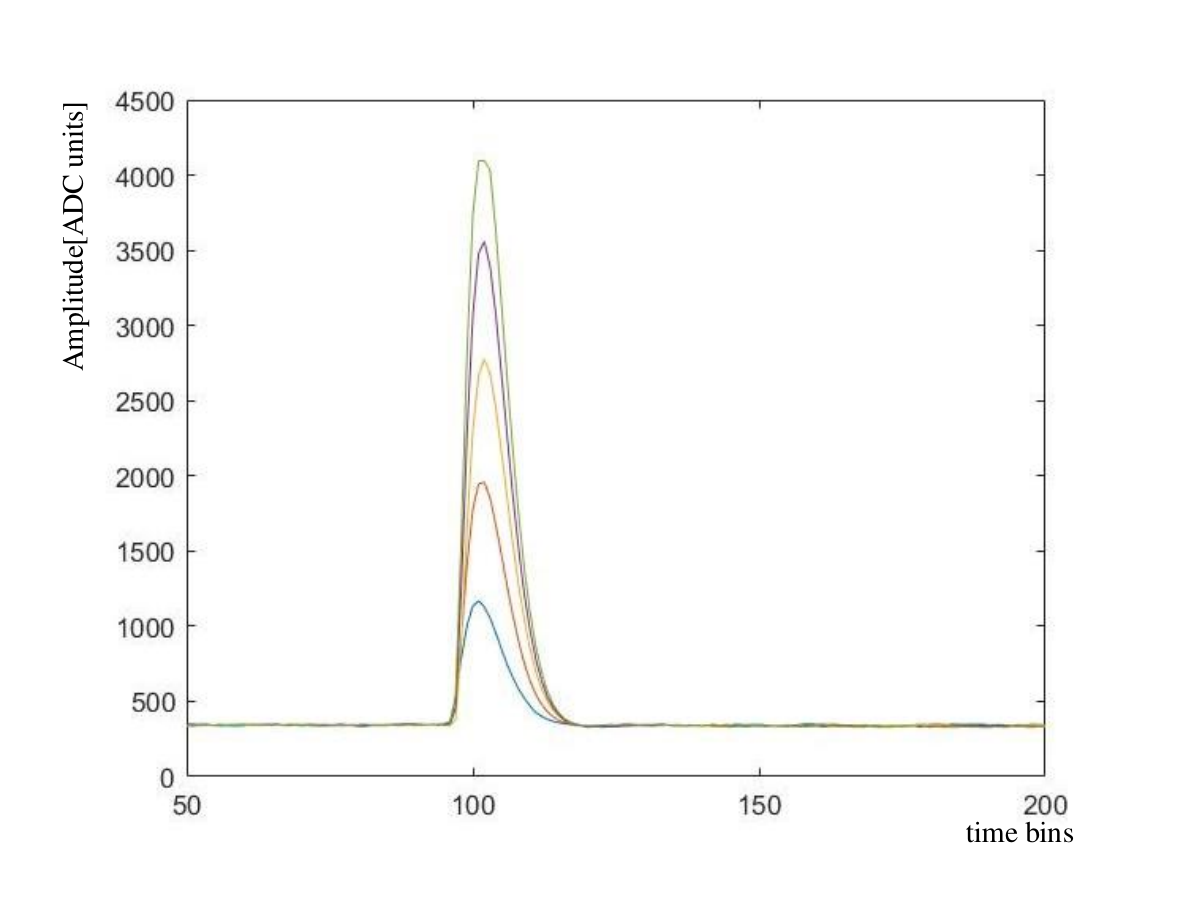}
  }
  \\
  \subfloat[INL test of electronics ]
  {
      \label{fig:subfig3}\includegraphics[width =55mm]{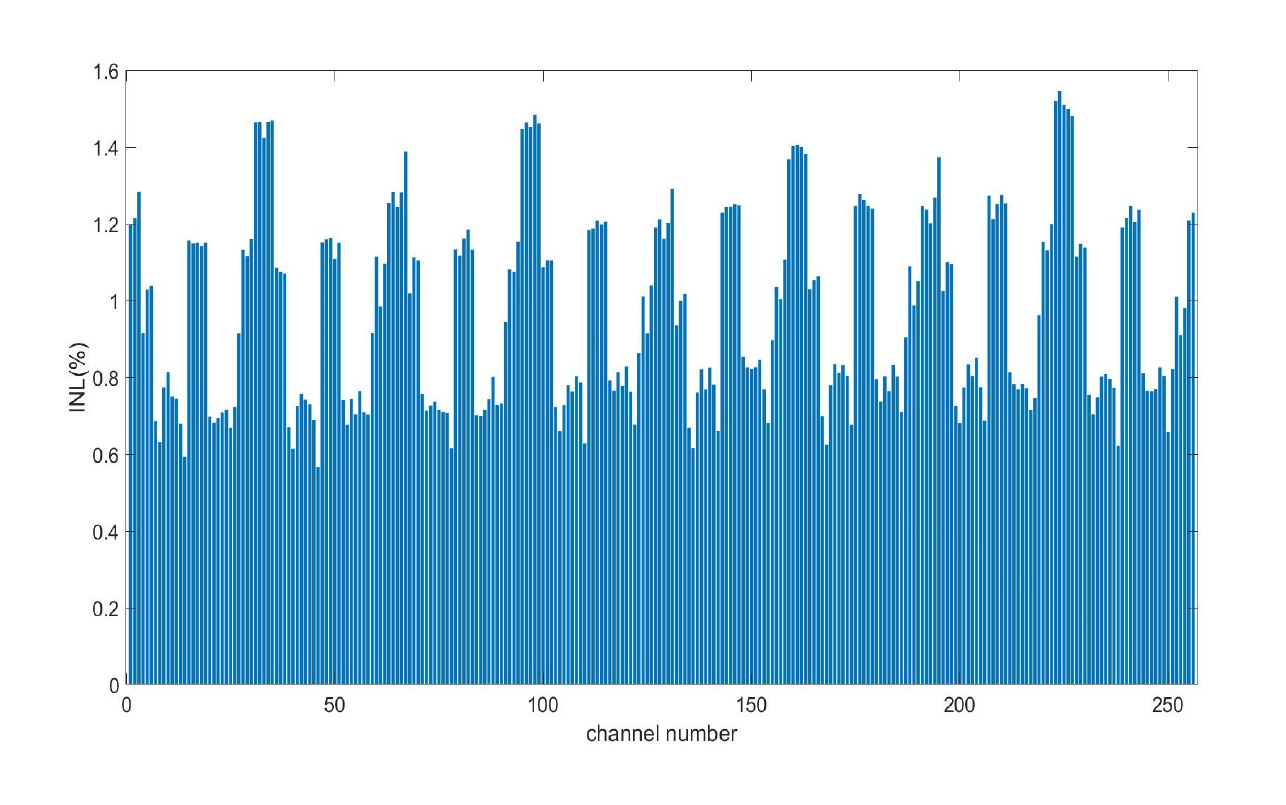}
  }
  \subfloat[test of data acquisition]
  {
      \label{fig:subfig4}\includegraphics[width =55mm]{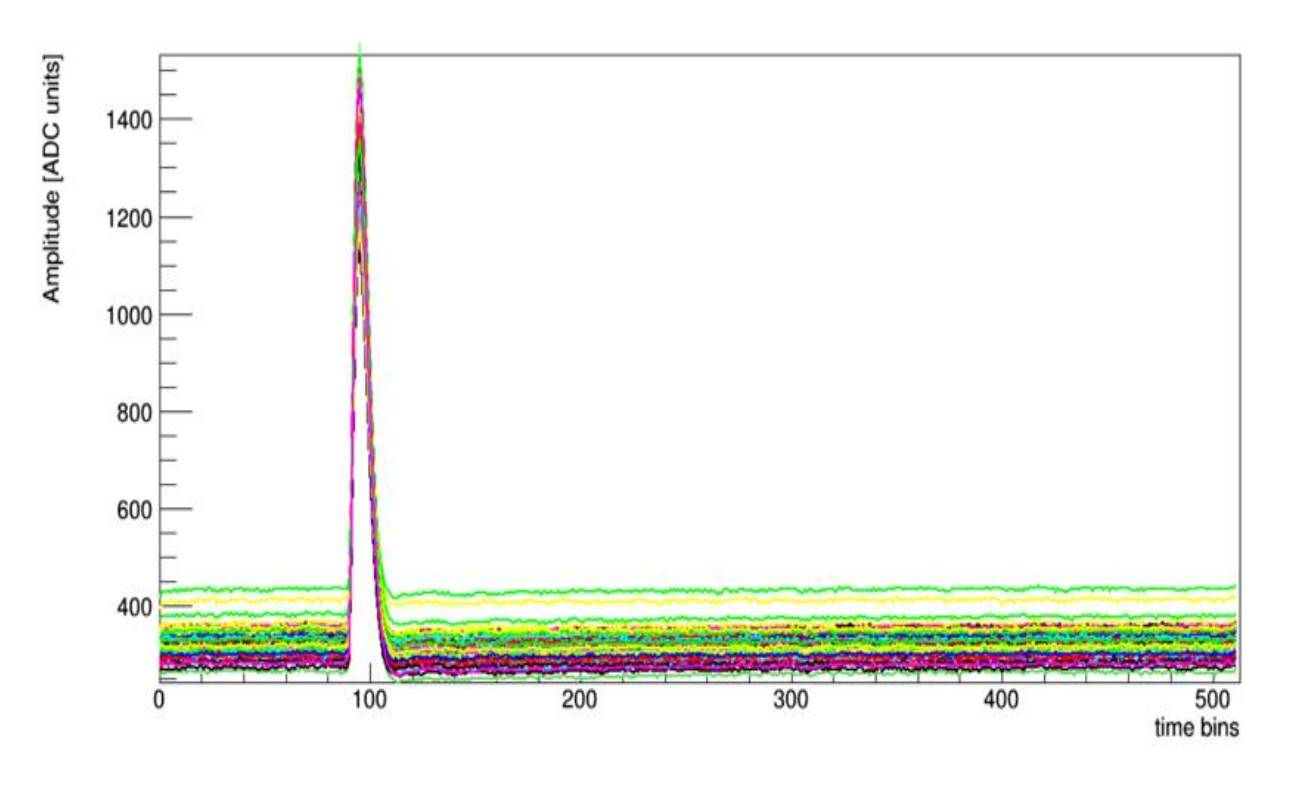}
  }
\caption{\label{fig:v} Result of the electronics test in PandaX-\uppercase\expandafter{\romannumeral3}. }
\end{figure}

The back-end electronics are connected to 26 front-end boards through normal IO optical fiber interfaces\cite{fiber}.  The noise test can visually show whether the electronics are normal. As can be seen in figure \ref{fig:subfig1}, The four noise maps correspond to the data collected by the four ASIC chips\cite{aget}\cite{aget2} on the front-end board. To compare noise levels between channels, we calculated the root mean square (RMS) of noise. each ASIC chip has four extra fixed pattern noise (FPN) channels, the noise levels of the other 64 channels are similar, and the operation is normal. In the self-scale test\cite{muography}\cite{aget2}, the back end sends configuration commands to the front end so that the ASIC chip on the front-end board generates a signal of corresponding amplitude. The test can simulate the actual operation of the experiment. As shown in figure \ref{fig:subfig2}, the back end can send data successfully to the front end. For the back-end electronics, the Integral non-linearity test of front-end board channels can show that the back-end can stably accept data from the front-end. The data acquisition test shows that the electronics can successfully capture signal instances. As shown in figure \ref{fig:v}, all the electronics tests are normal, and the back-end electronics are suitable for  PandaX-\uppercase\expandafter{\romannumeral3} experiments. Currently, it is being applied in  PandaX-\uppercase\expandafter{\romannumeral3}.

\subsection{Apply in VLAST}
\begin{figure}[htbp]
\centering 
\includegraphics[scale =0.22]{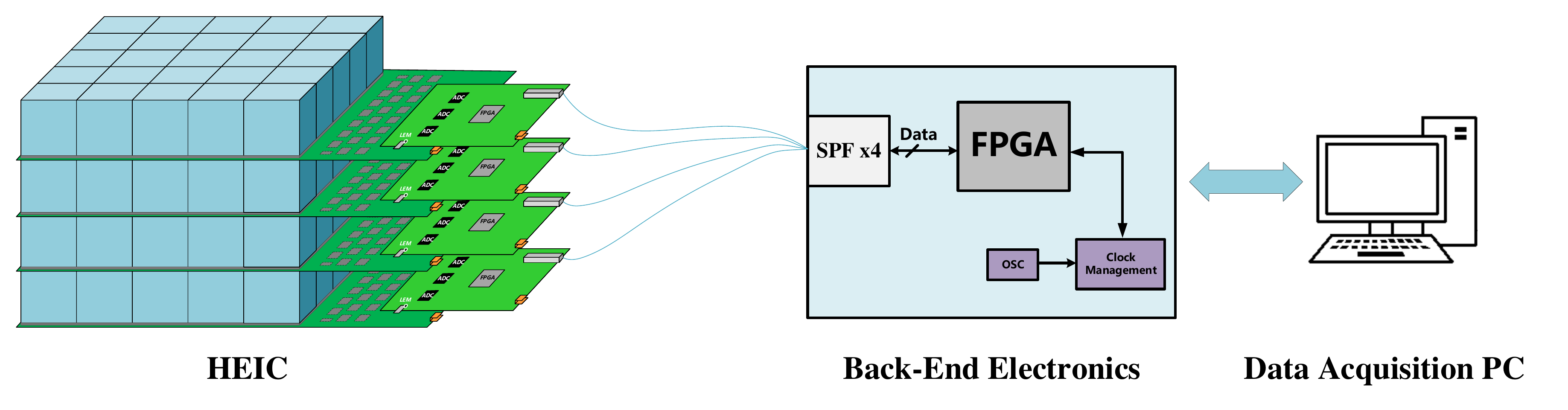}
\caption{\label{fig:z} Schematic of electronics test in VLAST. }
\end{figure}
\begin{figure}[htbp]
\centering 

\subfloat[beam test of muons]
  {
      \label{fig:subfig5}\includegraphics[width =45mm]{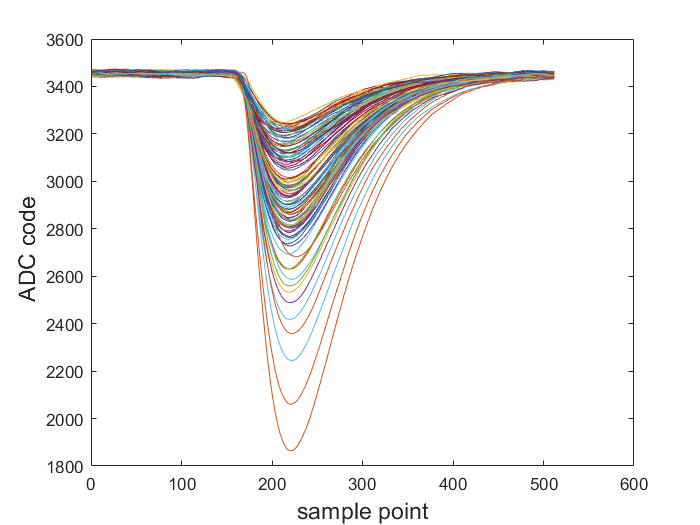}
  }
\subfloat[single channel of muons]{
    \label{fig:subfig6}\includegraphics[width =45mm]{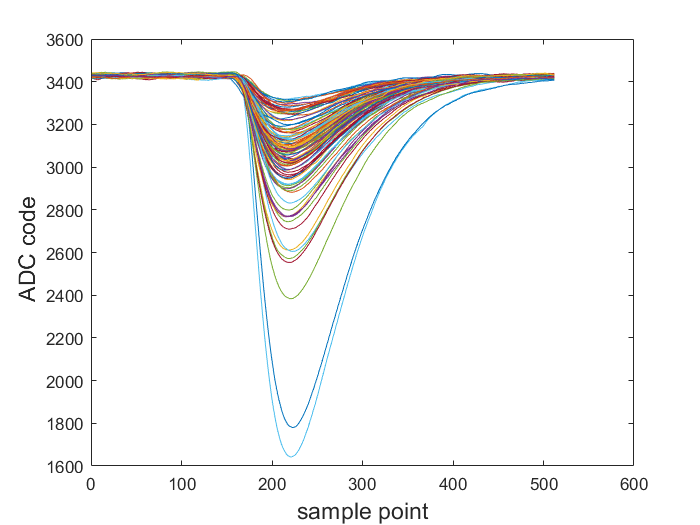}
}  
\\
\subfloat[beam test of electrons]
  {
      \label{fig:subfig7}\includegraphics[width=45mm]{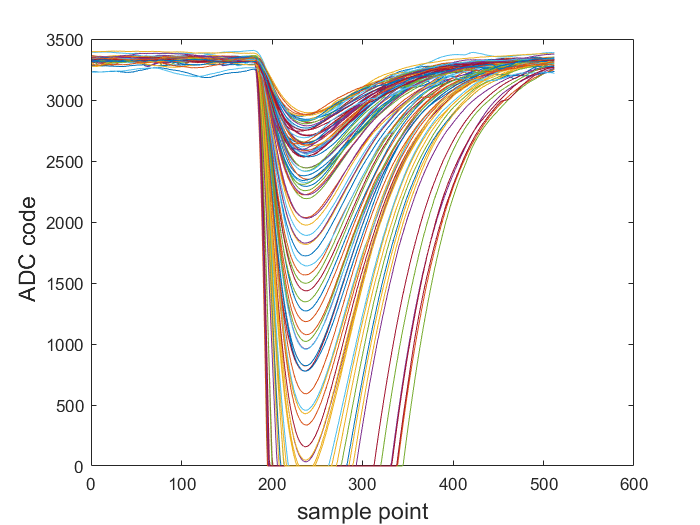}
  }
  \subfloat[single channel of electrons]{
    \label{fig:subfig8}\includegraphics[width=45mm]{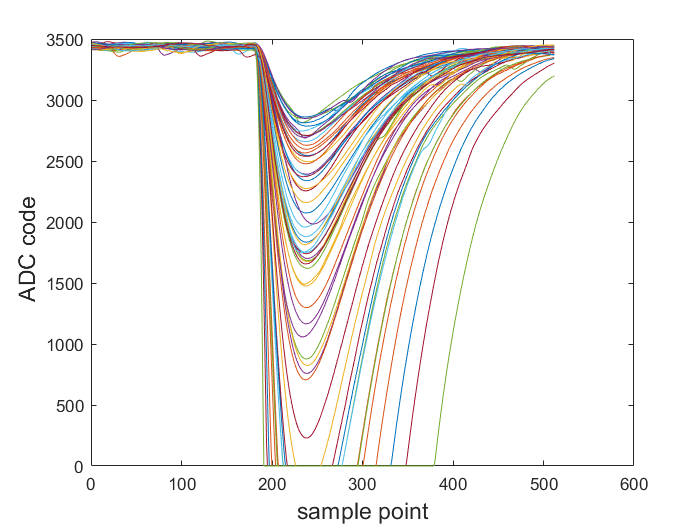}
} 
\caption{\label{fig:p} Result of the beam test in VLAST. }

\end{figure}

The detector prototype of the VLAST project underwent beam testing in the PS and SPS beam facilities at CERN in September 2023. This back-end electronics was applied in the sub-detector HEIC. In the electronics of HEIC, the data rate of the four front-end electronics is approximately 2.0Gbps. The front-end and back-end communication uses GTX protocol worked at 2.4Gbps\cite{chen_2022}.In the beam experiment, this set of electronics participated in the beam test of muons and electrons. The test results are shown in the figure\ref{fig:p}. The Back-end electronics work stably and normally. Analysis of experimental data is still in progress.

\subsection{Apply in Muongraphy}
The cosmic ray Muongraphy is of great significance for nuclear radiation, nuclear material detection, and non-destructive imaging of large-scale objects\cite{pyramids0}\cite{pyramids1}\cite{pyramids2}. It is also a popular research direction in current particle imaging technology. The team of the University of Science and Technology of China researched the readout electronics of high-resolution muon imaging systems based on Micromegas detectors and developed a set of experimental devices that meet the requirements of scattering imaging and transmission imaging called $\mu$STC( muon Scattering and Transmission imaging facility)\cite{wangyu}. The set of back-end electronics has been successfully used in the readout electronics of this experimental device. Similar to PandaX-\uppercase\expandafter{\romannumeral3}, the back-end electronics communicate with the front-end via common IO optical fiber interfaces.
\begin{figure}[htbp]
\centering 
\includegraphics[scale =0.4]{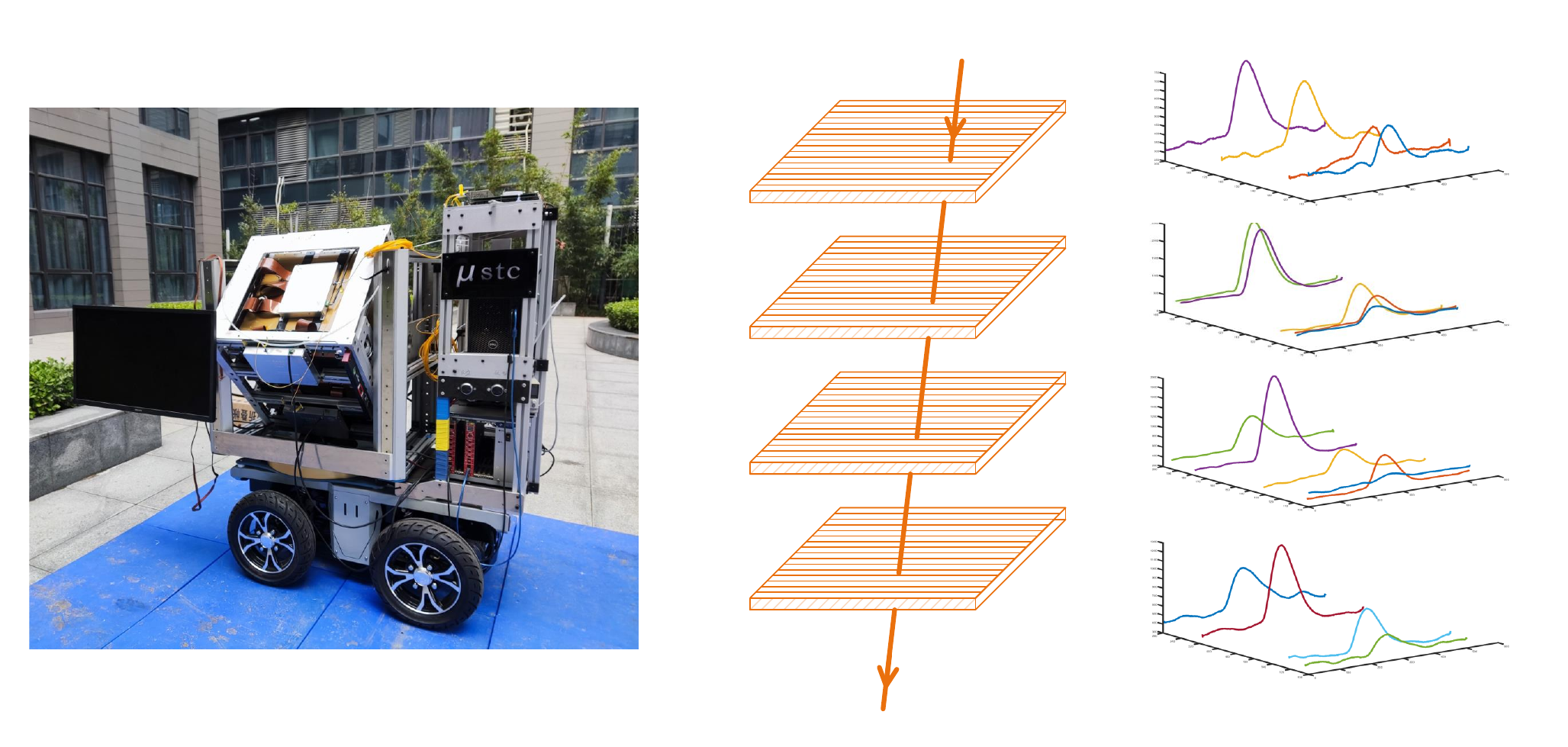}
\caption{\label{fig:s} Picture of moungraphy. }
\end{figure}

Figure \ref{fig:s} shows the facility of $\mu$STC and the schematic diagram of Micromegas. The back-end electronics are located on the upper right side of the facility. Researchers decoded where the muons hit based on signals collected by the back-end electronics.

\section{Summary}
This article introduces a portable back-end electronics design scheme with good versatility for small and medium-sized physics experiments. It adopts a daughter-motherboard design, allowing electronics to have greater flexibility. It supports 32 front-end boards based on the fibre protocol implemented by normal IO, and each front-end board has a bandwidth of around 200Mbps. Moreover, it supports 16 front-end boards based on GTX protocol, and each board has a bandwidth of several Gbps. It integrates TLU functions to serve physical experiments better and provide convenience for researchers. Due to the success of the design, this set of back-end electronics has been used in experiments such as  PandaX-\uppercase\expandafter{\romannumeral3}, VLAST, and Moungraphy.

\acknowledgments

This work was supported in part by National Science Foundation for Distinguished Young Scholars (Grant No.: 12025504), in part by National Natural Science Foundation of China (Grant No.: 12205297). The authors would like to thank Zhiyong Zhang from $\mu$STC group and Shuaifeng Guo for assistance with detector joint test. We also thank  PandaX-\uppercase\expandafter{\romannumeral3} collaboration for their help to this work.


\end{document}